\documentclass[showpacs,12pt,a4paper]{iopart}
\usepackage{amssymb}
\usepackage{amsfonts}

\def\<{\langle}
\def\>{\rangle}
\def\H{{\cal H}}
\def\N{{\mathbb N}}
\def\C{{\mathbb C\, }}

\def\PP{{\mathbb P}}
\def\xx{{\bf x}}
\def\yy{{\bf y}}
\def\zz{{\bf z}}
\def\tr{{\rm tr\,}}

\begin{document}

\title[Covariants of four qubits]{A complete set of covariants of the four qubit system}

\author{Emmanuel Briand, Jean-Gabriel Luque and  Jean-Yves Thibon}

\address{Institut Gaspard Monge, Universit\'e de Marne-la-Vall\'ee\\
77454 Marne-la-Vall\'ee cedex, France}

\date{}

\begin{abstract}
We obtain a complete and minimal set of 170 generators for
the algebra of $SL(2,\C)^{\times 4}$-covariants
of a binary quadrilinear form. 
Interpreted in terms of a four qubit system, this
describes  in particular the algebraic varieties 
formed by the orbits of  local filtering
operations in its projective Hilbert space.
Also, this
sheds some light on the   local unitary
invariants, and provides all the possible
building blocks for the construction of entanglement measures
for such a system.
\end{abstract}



\section{\label{intro}Introduction}

The invariant theory of fourth rank hypermatrices (or
quadrilinear forms) over a two-dimensional complex vector space,
already considered at the end of the nineteenth century, 
has experienced
recently a regain of interest, mainly due to its potential applications
to the understanding of entanglement in four qubit systems
\cite{Tap,Mi,Ver,LT1}. We shall give here the first complete
solution to the problem of describing the polynomial covariants,
an investigation started by Le Paige as early as 1881
\cite{LeP1}.
The theory was further advanced by C Segre in 1922 \cite{Segre},
using only geometric methods  which led him close to a complete
classification of the orbits. Such a complete classification
was obtained only recently by Verstratete et al. \cite{Ver},
by exploiting the local isomorphism between $SO_4$ and
$SL_2\times SL_2$, which permits a reduction of the problem
to the classification of complex symmetric matrices up
to orthogonal transformations.

A complete picture would require a description of the
orbit closures as algebraic varieties and an understanding
of their ordering with respect to inclusion. 
In classical invariant theory, such descriptions were
usually given in terms of invariants and covariants.
The main result of \cite{LT1} was a complete
description of the algebra of  polynomial functions $f(a_{ijkl})$
in the components of a four qubit state
\begin{equation}
 |\Psi\>=\sum_{i,j,k,l=0}^1 a_{ijkl} |i\>\otimes |j\>\otimes |k\>\otimes |l\>
\end{equation}
which are invariant under
the natural action of the 
so-called SLOCC\footnote{for {\it Stochastic Local Operations assisted
by Classical Communication}, see  \cite{Dur}.}
group $G=SL(2,\C)^{\times 4}$ on the
local Hilbert space  $\H=V^{\otimes 4}$, where $V=\C^2$.

One motivation for this investigation was to test on
the four qubit case Klyachko's proposed definitions of entanglement
and complete entanglement \cite{Klyachko}. These consist
in identifying entangled states as being precisely those for which
at least one SLOCC polynomial invariant is not zero, and completely
entangled states as the vectors of minimal norm in closed SLOCC orbits,
which are unique up to local unitary transformations.

Klyachko's definition of complete entanglement seems to be supported
by the recent numerical experiments of Verstraete et al.
\cite{Ver2}. Indeed, these authors propose a numerical
algorithm converging to a normal form, which, in the case
of a stable state, is a state of minimal norm in its
SLOCC orbit ${\cal O}$, and otherwise in the unique closed orbit contained
in the  closure $\bar{\cal O}$. Thus, in both cases, the normal form
is a completely entangled state in the sense of
\cite{Klyachko}.

We take the opportunity to point out that,
as conjectured in \cite{Ver2},
the normal
form is indeed
unique up to local unitary transformations
\footnote{
The results implying this conjecture, as
well as the necessary background in Geometric Invariant Theory,
are collected in
\cite{Encycl}: the Kempf-Ness criterion (theorem 6.18)
proves the result in the case of a closed orbit, and by the corollary
of theorem 4.7, there is a unique closed orbit in the closure of
an arbitrary orbit.}.

In \cite{LT1}, the polynomial invariants were constructed by means
of the classical notion of a covariant. If we
interpret our state $|\Psi\>$ as a quadrilinear form
\begin{equation}
f(\xx,\yy,\zz,{\bf t})=\sum_{i,j,k,l=0}^1a_{ijkl}x_iy_jz_kt_l
\end{equation}
on $V\times V\times V\times V$, a covariant of $A$ is a
multi- homogeneous $G$-invariant polynomial
in the form coefficients $a_{ijkl}$ and in the original
variables $x_i,y_j,z_k,t_l$.

Since the spaces $S^\mu(V)$ ($\mu\in\N^4$) of homogeneous
polynomials of multidegree $\mu$ in $\xx,\yy,\zz,{\bf t}$ exhaust
all finite dimensional representations of $G$, a covariant
of degree $d$ in the $a_{ijkl}$ and $\mu$ in the variables
can be regarded as a $G$-equivariant map $S^d(\H)\rightarrow S^\mu(V)$
from the space $S^d(\H)$ of homogeneous polynomials of degree $d$
in $A$ to the irreducible representation $ S^\mu(V)$.
Such a map is determined by the image of a highest weight vector,
so that covariants are in one to one correspondence with
highest weight vectors in $S^d(\H)$, these being known
as semi-invariants in the classical language (cf. \cite{Olver}).

The covariants form an algebra, which is naturally graded
with respect to $d$ and $\mu$.
We denote by  ${\cal C}_{d;\mu}$ the corresponding graded
pieces.
The knowledge of their dimensions $c_{d;\mu}$
is equivalent to the decomposition of the character
of $S^d(\H)$ into irreducible characters of $G$,
and the knowledge of a basis of ${\cal C}_{d;\mu}$
allows one to write down a Clebsch-Gordan series
with respect to $G$
for any polynomial in the $a_{ijkl}$.
Also, it is known that the
equations of any $G$-invariant closed subvariety
of the projective space $\PP(\H)$ are given by the
simultaneous vanishing of the coefficients of some
covariants.

Finally, let us point out a connection with
the approach of \cite{GBRM} and local unitary
invariants.
The   spaces $S^\mu(V)$, and hence the ${\cal C}_{d;\mu}$
are also Hilbert spaces in a natural way. If
$\Psi_{d;\mu}^{\alpha}$ is a linear basis of  ${\cal C}_{d;\mu}$,
the scalar products $\<\Psi_{d;\mu}^{\alpha}|\Psi_{d;\mu}^{\beta}\>$
(taken with respect to the variables, the coefficients $a_{ijkl}$
being treated as scalars) form a basis of the space of
$U(2)^{\times 4}$ invariants of degree $2d$ (that is, $d$ in $A$ and
$d$ in $\bar A$),
and the $\<\Psi_{d;\mu}^{\alpha}|\Psi_{e;\mu}^{\beta}\>$
form a basis of the space of $SU(2)^{\times 4}$ invariants of
bidegree $(d,e)$ in $(A,\bar A)$. Such expressions are used
for example in  \cite{Dur}, in the case of three qubits.

\section{Summary of method and results}

A minimal generating set consisting of 170 covariants
is found by means a computer search through {\it
iterated transvectants} (see section \ref{transv}),
guided by the knowledge of the
Hilbert series (see section \ref{hilbert}),
and simplified by taking into account
some special properties
of multilinear forms.
The following table gives the number of covariants
of degree $d$ in $A$ and multidegree $\lambda$ in the variables,
where $\lambda$ is in nondecreasing order. There are
similar covariants for each of the $n_\lambda$ permutations
$\mu$ of the degrees.
For example, in degree 5, we have 12 covariants which are
cubic in one variable and linear in the other three, and one
quadrilinear covariant.

\begin{equation}
\begin{array}{|c||c||c|c|c|c|c|c|c|c|c|c|c|c|}
\hline\lambda\backslash d&n_\lambda&1&2&3&4&5&6&7&8&9&10&11&12\\
\hline  0000&1&&1&&2&&1&&&&&& \\
\hline 1111&1&1&&2&&1&&&&&&&\\
\hline 2200&6&&1&&1&&1&&&&&&\\
\hline 2220&4&&&&2&&2&&2&&&&\\
\hline 3111&4&&&1&&3&&3&&1&&&\\
\hline 3311&6&&&&&&&1&&2&&1&\\
\hline 4000&4&&&&1&&&&1&&&&\\
\hline 4200&12&&&&&&1&&1&&1&&\\
\hline  5111&4&&&&&&&1&&2&&1&\\
\hline 6000&4&&&&&&&&&&&&1\\
\hline
\end{array}
\nonumber
\end{equation}

\section{\label{transv}Multiple transvectants}

Transvectants, or Cayley's Omega-process, are the basic tools for constructing
complete systems of covariants, and play a key r\^ole in Gordan's
and Hilbert's proofs that the ring of covariants is finitely generated.
The notion of a transvectant extends with little modifications to
forms
in several series of variables, and appears to have been first
exploited by
Le Paige \cite{LeP} in the case of binary
trilinear forms, and by
Peano \cite{Pea}, who computed the complete systems for
forms of bidegrees $(1,1)$, $(2,1)$ and $(2,2)$ in two independent
binary variables. Complete systems for bidegrees $(3,1)$ and $(4,1)$
have been given by Todd \cite{To1,To2}, and, to the best of our knowledge,
the only forms in more that two binary variables for which the complete
system is known are the $(1,1,1)$ \cite{LeP,Sad,Sw} and the $(2,1,1)$, due
to Gilham \cite{Gil3}. The geometry of the quadrilinear form is
discussed by Segre \cite{Segre} but no attempt is made to describe the
covariants.

If $f$ and $g$ are forms in the binary variable $\xx=(x_1,x_2)$,
we identify their tensor product $f\otimes g$ with the polynomial
$f(\xx')g(\xx'')$ in two independent binary variables $\xx'$, $\xx''$.
Following \cite{Olver}, the multiplication map $f\otimes g\mapsto fg$
is denoted by $\tr$. So, $\tr(f(\xx')g(\xx''))=f(\xx)g(\xx)$.

The Cayley operator $\Omega_\xx$ acts on such a tensor product
by the differential operator
\begin{equation}
\Omega_\xx=\left|
\matrix{
\frac{\partial}{\partial x'_1} & \frac{\partial}{\partial x''_1} \cr
\frac{\partial}{\partial x'_2} & \frac{\partial}{\partial x''_2}
}\right|
\end{equation}

If $f$ and $g$ are two $p$-tuple forms in $p$ independent
binary variables $\xx_i$, one
defines for any $(i_1,i_2,\ldots,i_p)\in\N^p$ a
multiple transvectant of $f$ and $g$ by
\begin{equation}
(f,g)^{i_1i_2\ldots i_p}=\tr \Omega_1^{i_1}\Omega_2^{i_2}\cdots \Omega_p^{i_p}
f(\xx_1',\ldots,\xx_p')g(\xx_1'',\ldots,\xx_p'')\,,
\end{equation}
where $\Omega_i=\Omega_{\xx_i}$, and $\tr$ acts on all variables
by $\xx'_i,\xx''_i\mapsto \xx_i$.

It can be proved that the complete system of covariants of any number
of forms can be reached in a finite number of steps by building
iterated transvectants, starting with the ground forms.

\section{\label{hilbert}The Hilbert series}

The (multivariate) Hilbert series for the algebra of covariants is 
defined by
\begin{equation}
h(t,u_1,u_2,u_3,u_4)=
\sum_{d,\mu} c_{d;\mu} t^d {\bf u}^{\mu}\,,
\end{equation}
where $c_{d;\mu}$ is the dimnension of the space of homogeneous
covariants which are of degree $d$ in the $a_{ijkl}$ and
of multidegree $\mu$ in the variables.
Let
\begin{equation}
S=\prod_{i=1}^4(1-u_i^{-2})\prod_{\alpha\in\{-1,1\}^4}(1-{\bf
u}^\alpha t)^{-1} \,.
\end{equation}
Here $S$ has to be considered as the formal power series obtained by
expansion with respect to the variable $t$.
Let ${\cal L}$ be the linear operator
acting on a formal series in $t,{\bf u}$ by leaving unchanged
every monomial $t^d {\bf u}^{\mu}$ with $\mu \in \N^4$,
and annihilating those with $u$-exponent having some negative coordinate.
It follows from standard considerations
about the characters of $G$ that $h={\cal L} S$.

By successive decompositions into partial fractions
(with respect to $u_1$, next $u_2,u_3,u_4$)
we have computed this series,
which guided us in the search for the covariants.
The numerator is too large to be printed, but if one substitutes
$u_1=u_2=u_3=u_4=u$, one finds after simplification $h=P/Q$, where
the numerator $P$ is
\begin{eqnarray*}
1-{u}^{2}t+ \left( 3\,{u}^{4}-2\,{u}^{2} \right) {t}^{2}+
 \left( {u}^{6}+4\,{u}^{4} \right) {t}^{3}
\nonumber
\\
+\left( 10\,{u}^{4}-{u}^{2} \right) {t}^{4}+ \left( -4\,{u}^{8}-2\,{u}^{6}+2\,{u}^{4}
 \right) {t}^{5}\nonumber\\
+ \left( 2\,{u}^{10}+6\,{u}^{8}-2\,{u}^{6}+8\,{u}^{4} \right) {t}^{6}+ \left( 2\,
{u}^{10}+6\,{u}^{8} \right) {t}^{7}\nonumber\\
+ \left( -8\,{u}^{12}+{u}^{10}+13\,{u}^{8}-2\,{u}^{6}+4\,{u}^{4} \right) {t}^{8}
\nonumber\\
+ \left( -8\,{u}^{12}-{u}^{10}+12\,{u}^{8}-{u}^{6} \right) {t}^{9}
\nonumber\\
+ \left( 2\,{ u}^{14}-13\,{u}^{12}+13\,{u}^{8}-2\,{u}^{6} \right) {t}^{10}
\nonumber\\
+ \left( {u}^{14}-12\,{u}^{12}+{u}^{10}+8\,{u}^{8} \right) {t}^{11}
\nonumber
\\
+ \left( -4\,{u}^{16}+2\,{u}^{14}-13\,{u}^{12}-{u}^{10}+8\,{u}^{8} \right){t}^{12}
\nonumber\\
+ \left( -6\,{u}^{12}-2\,{u}^{10} \right) {t}^{13}+
\left( -8\,{u}^{16}+2\,{u}^{ 14}-6\,{u}^{12}-2\,{u}^{10} \right) {t}^{14}
\nonumber\\
+ \left( -2\,{u}^{16}+2\,{u}^{14}+4\,{u}^{12} \right) {t}^{15}
+ \left( {u}^{18}-10\,{u}^{16} \right) {t}^{16}
\nonumber\\
+ \left( -4\,{u}^{16}-{u}^{14}
 \right) {t}^{17}+ \left( 2\,{u}^{18}-3\,{u}^{16} \right) {t}^{18}+{u}^{18}{t}^{19}-{u}^{20}{t}^{
20}
\end{eqnarray*}
and the denominator $Q$ is
\begin{eqnarray*}
(1-tu^2)(1-t u^4)(1-t^2)(1-t^2u^2)^2(1-t^2u^4)^3\\ \times
(1-t^4)(1-t^4 u^2)(1-t^4u^4)(1-t^6)\,.
\end{eqnarray*}
The algebra of covariants is Cohen-Macaulay (see, e.g., \cite{Encycl}).
This means that it is a free module of finite rank over a subalgebra
generated by a finite familly of homogeneous, algebraically independent elements $f_1,\ldots,f_{k}$.
Then $k$ is the Krull dimension of the algebra of covariants
(the maximum number of algebraically independent elements)
and $h(t,t,t,t,t)$ has a pole at $t=1$, of order $k$. We found in this way that $k=12$.

When substituting  $u_i=0$, the Hilbert series of the invariants is recovered.

\section{\label{covariants}A fundamental set of covariants}

The fundamental covariants are tabulated in  \ref{tabcov}.
They are denoted by symbols $X_{pqrs}^{m}$,
in which the letter $X$ indicates the degree in the
form coefficients $a_{ijkl}$ (1 for $A$, 2 for $B$, etc.),
the subscripts $ijkl$ indicate the degrees in the variables
$\xx,\yy,\zz,{\bf t}$, and the optional exponent $m$ serves
to distinguish between covariants having the same degrees.

The only covariant of degree $1$ is the ground form $f=A_{1111}$.
In degree two, we have one invariant, the basic hyperdeterminant
$H=\case12 B_{0000}$, which is defined in general for all
multilinear form with an even number of variables
(see \cite{LT0} for some applications), and six biquadratic
forms such as $B_{2200}$. In degree 3, we find two quadrilinear
forms and four cubico-trilinear covariants. There are
three other invariants, which occur in degree 4 ($D^1_{0000}$, $D^2_{0000}$)
and 6 ($F_{0000}$).
The maximal degree is 12, where we find four binary sextics.

We see that some of the covariants are of a type for which
the invariant theory is well understood (binary forms of degree
4 and 6, biquadratic forms), or are again quadrilinear forms.
Other are of essentially unexplored types (3111, 2220, 4200, 3311, 5111).

Naturally, the covariants of a covariant are again covariants
(although not necessarily irreducible). Some of these are discussed
in  \ref{invcov}.

Let us now sketch the application to the description
of the orbit closures.
The normal forms obtained by Verstraete et al. \cite{Ver} are

\begin{eqnarray*}
G_{abcd}&=&\frac{a+d}{2}(|0000\rangle
+|1111\rangle)+\frac{a-d}{2}(|0011\rangle
+|1100\rangle)\\
&&\hspace{.1cm}+\frac{b+c}{2}(|0101\rangle
+|1010\rangle)+\frac{b-c}{2}(|0110\rangle +|1001\rangle)\\
L_{abc_2}&=&\frac{a+b}{2}(|0000\rangle
+|1111\rangle)+\frac{a-b}{2}(|0011\rangle +|1100\rangle)\\
&&\hspace{.5cm}+c(|0101\rangle +|1010\rangle)+|0110\rangle \\
L_{a_2b_2}&=&a(|0000\rangle +|1111\rangle)+b(|0101\rangle+|1010\rangle)\\
&&\hspace{.5cm}+|0110\rangle +|0011\rangle \\
L_{ab_3}&=&a(|0000\rangle
+|1111\rangle)+\frac{a+b}{2}(|0101\rangle
+|1010\rangle)\\
&&\hspace{.5cm}+\frac{a-b}{2}(|0110\rangle +|1001\rangle)\\
&&\hspace{.5cm}+\frac{i}{\sqrt{2}}(|0001\rangle +|0010\rangle
+|0111\rangle
+|1011\rangle)\\
L_{a_4}&=&a(|0000\rangle +|0101\rangle +|1010\rangle
+|1111\rangle)\\
&&\hspace{.5cm}+(i|0001\rangle +|0110\rangle -i|1011\rangle)\\
L_{a_20_{3\oplus\bar{1}}}&=&a(|0000\rangle
+|1111\rangle)+(|0011\rangle +|0101\rangle +|0110\rangle)\\
L_{0_{5\oplus\bar{3}}}&=&|0000\rangle +|0101\rangle +|1000\rangle
+|1110\rangle \\
L_{0_{7\oplus\bar{1}}}&=&|0000\rangle +|1011\rangle +|1101\rangle
+|1110\rangle\\
L_{0_{3\oplus\bar{1}}0_{3\oplus\bar{1}}}&=&|0000\rangle+|0111\rangle
\end{eqnarray*}                                                            
where in our notation a ket $|ijkl\rangle$ is to be identified with the
monomial $x_iy_jz_kt_l$.              
The invariants $B_{0000}$, $D^1_{0000}$, $D^2_{0000}$ and $F_{0000}$ are shown in
\cite{LT1} to
separate the normal forms $G_{abcd}$, $L_{abc_2}$, $L_{ab_3}$, $L_{a_2b_2}$,
$L_{a_4}$ and $L_{a_20_{3\oplus\overline1}}$. But they
vanish for $L_{0_{5\oplus\overline 3}}$, $L_{0_{7\oplus\overline 1}}$ and
$L_{0_{3\oplus\overline1}0_{3\oplus\overline1}}$. The knowledge of the
fundamental set of covariants is more than sufficient to separate the last three
forms. Indeed,
\begin{eqnarray}
C_{3111}(L_{0_{5\oplus\overline
3}})=&2(x_2y_2z_1t_1-x_1y_2z_1t_1)\\D_{2200}(L_{0_{5\oplus\overline 3}})=&0\\
C_{3111}(L_{0_{7\oplus\overline
1}})=&2x_2(y_1z_1t_2+y_1z_2t_1-2y_2z_1t_1)\\D_{2200}(L_{0_{7\oplus\overline 1}})=&-16x_2^2z_1z_2\\
C_{3111}(L_{0_{3\oplus\overline
1}0_{3\oplus\overline
1}})=&0\\D_{2200}(L_{0_{3\oplus\overline
1}0_{3\oplus\overline
1}})=&0
\end{eqnarray}
From this, it is not difficult (although somewhat tedious)
to obtain the equations of the
closures of the orbits, and to study the inclusions between them.
This will be left for a future paper on the subjet, in which
we expect to be able to produce first a better choice of the fundamental
covariants (with greater geometrical or physical significance).

\section{\label{ratcov}Rational covariants}

The  algebra of  rational covariants is simpler
than the algebra of polynomial covariants. It is a field
of rational functions over 12 homogeneous independent generators.
A way to compute a
fundamental set of rational semi-invariants, consists in using
the so-called
associated forms \cite{Me}. Let
$F$ be the polynomial obtained from the ground form by
applying the following series of substitutions
\begin{eqnarray}
x_1&\rightarrow a_{0000}x_1-a_{1000}x_2,\,& x_2\rightarrow
a_{0000}x_2,\nonumber\\
y_1&\rightarrow a_{0000}y_1-a_{0100}y_2,\,&
y_2\rightarrow a_{0000}y_2,\nonumber\\
z_1&\rightarrow
a_{0000}z_1-a_{0010}z_2,\,& z_2\rightarrow
a_{0000}z_2,\nonumber\\
t_1&\rightarrow a_{0000}t_1-a_{0001}t_2,\,&
t_2\rightarrow a_{0000}t_2.\nonumber
\end{eqnarray}
The semi-invariants which are the sources of the associated forms
are the coefficients of the monomials $x_iy_jz_kt_l$ in $F$,
divided by $a_{0000}^{9-i-j-k-l}$ 
We obtain in this way a list of semi-invariants which are the sources of some polynomial
covariants given below. Here, $H=\frac12(f,f)^{1111}$,
$b_{xy}=\frac12(f,f)^{0011}$, etc. are as in \cite{LT1}.
\begin{equation}
\begin{array}{|c|c|}
\hline 
\mbox{Source $c_{\alpha}$}&\mbox{Covariant ${\cal
C}_\alpha$}\\ 
\hline 
c_{0000}&1\\
c_{1000}&0\\
c_{0100}&0\\
c_{0010}&0\\
c_{0001}&0\\ 
c_{0011}&b_{xy}\\
c_{0101}&b_{xz}\\ 
c_{0110}&b_{xt}\\ 
c_{1001}&b_{yz}\\
c_{1010}&b_{yt}\\ 
c_{1100}&b_{zt}\\ 
c_{0111}&-C_{3111}\\
c_{1011}&-C_{1311}\\ 
c_{1101}&-C_{1131}\\ 
c_{1110}&-C_{1113}\\
c_{1111}&Hf^2-b_{xy}b_{zt}-b_{xz}b_{yt}-b_{xt}b_{yz}\\
\hline
\end{array}\nonumber
\end{equation}
The $12$ homogeneous independent generators of the field of rational covariants 
are the $11$ non-trivial associated forms above, together with the ground form $f$.
Actually, the last one (${\cal C}_{1111}$) can be advantageously replaced by $H$.

Now, each covariant can be written as a rational function in these $12$ generators. 
It suffices to make the substitutions
\begin{equation}\label{syz}
a_{ijkl}\rightarrow  
{\cal C}_{ijkl}  f^{1-(i+j+k+l)}
\end{equation}
 in the
source of the covariant, where ${\cal C}_\alpha$ is the covariant
with source  $c_\alpha$.

For example, the source of  $D_{4000}$ is
\begin{eqnarray}
2a_{0111}a_{0100}a_{0000}a_{0011}-4a_{0111}a_{0010}a_{0001}a_{0010}\nonumber\\-
a_{0000}^2a_{0111}^2+2a_{0111}a_{0000}a_{0001}a_{0110}\nonumber\\+2a_{0111}a_{0000}a_{0010}a_{0101}
+2a_{0110}a_{0001}a_{0011}a_{0100}\nonumber\\-a_{0001}^2a_{0110}^2+2a_{0110}a_{0001}a_{0010}a_{0101}
\nonumber\\-4a_{0110}a_{0011}a_{0000}
a_{0101}+2a_{0101}a_{0010}a_{0011}a_{0100}\nonumber\\-a_{0010}^2a_{0101}^2-a_{0011}^2a_{0100}^2\nonumber
\end{eqnarray}
and the above substitutions give
\begin{eqnarray}
D_{4000}&=-\frac1{{\cal C}_{0000}^2}({\cal C}_{0111}^2+4{\cal
C}_{0110}{\cal C}_{0011}{\cal C}_{0101})\nonumber\\
&=-\frac1{f^2}(C_{3111}^2+4b_{xt}b_{xy}b_{xz}).
\end{eqnarray}
This yields a syzygy
\begin{equation}
f^2D_{4000}+C_{3111}^2+4b_{xy}b_{xz}b_{xt}=0\,.
\end{equation}

\section{\label{concl}Conclusion}

It is remarkable that the investigation of the fine structure
of the four qubit system has led to the first complete solution
of a mathematical problem which had already been considered as early
as 1881 \cite{LeP1}. This problem was among the very few ones which were
out of reach of the computational skills of the classical invariant
theorists, though accessible to a computer treatment.
The number of fundamental covariants, here 170, is not, however,
the highest ever found \footnote{
For example, Turnbull obtained in 1910 a system of 784 forms
for the case of three ternary quadratics, and in 1947,
Todd proved that 603 of them formed a complete minimal system.},
and we expect to be able to produce in the near future
a human readable proof, together with a better choice
of the generators, i.e., to find, at least for the lowest degrees,
generators with a transparent geometrical interpretation.

A complete description of the ring of covariants should in principle
include a generating set of the syzygies. However, we can see
from the Hilbert series that this is a hopeless task, as it is
already for the previously known specializations.
We have computed all the syzygies up to degree 7, and
formula (\ref{syz}) allows one to find at least one syzygy
for each covariant which is not one of the ${\cal C}_\alpha$.

Turning back to the issue of entanglement, we see that we have
now at our disposal all the possible building blocks for
the construction of entanglement
measures for systems with no more than
four qubits. It is to be expected that further investigations
will allow one to select among them the most relevant ones,
and that the analysis of their geometric significance will
give a clue for the general case.


\appendix

\newpage
\section{\label{tabcov}Fundamental covariants}

\small
\begin{equation*}\begin{array}{cc}\mbox{Degree 2}&\mbox{Degree 3}\\
\begin{array}{|c|c|}
\hline \mbox{Symbol}&\mbox{Transvectant}\\
\hline
B_{0000}&(f,f)^{1111}\\
\hline B_{2200}&(f,f)^{0011}\\
B_{2020}&(f,f)^{0101}\\
B_{2002}&(f,f)^{0110}\\
B_{0220}&(f,f)^{1001}\\
B_{0202}&(f,f)^{1010}\\
B_{0022}&(f,f)^{1100}\\\hline
\end{array}\nonumber&\begin{array}{|c|c|}
\hline \mbox{Symbol}&\mbox{Transvectant}\\
\hline C^1_{1111}&(f,B_{2200})^{1100}\\
 C^2_{1111}&(f,B_{2020})^{1010}\\\hline C_{3111}&(f,B_{2200})^{0100}\\
 C_{1311}&(f,B_{2200})^{1000}\\
 C_{1131}&(f,B_{2020})^{1000}\\
 C_{1113}&(f,B_{2002})^{1000}\\
 \hline
\end{array}\nonumber \end{array}
\end{equation*}
\begin{center}Degree 4
\begin{equation*}\begin{array}{cc}\begin{array}{|c|c|}\hline
\mbox{Symbol}&\mbox{Transvectant}\\\hline
D^1_{0000}&(f,C^1_{1111})^{1111}\\
D^2_{0000}&(f,C^2_{1111})^{1111}\\\hline
D_{2200}&(f,C_{3111})^{1011}\\
D_{2020}&(f,C^1_{1111})^{0101}\\
D_{2002}&(f,C_{3111})^{1110}\\
D_{0220}&(f,C_{1311})^{1101}\\
D_{0202}&(f,C_{1311})^{1110}\\
D_{0022}&(f,C_{1131})^{1110}\\\hline
\end{array}&\begin{array}{|c|c|}\hline
\mbox{Symbol}&\mbox{Transvectant}\\\hline
D_{4000}&(f,C_{3111})^{0111}\\
D_{0400}&(f,C_{1311})^{1011}\\
D_{0040}&(f,C_{1131})^{1101}\\
D_{0004}&(f,C_{1113})^{1110}\\\hline
D^1_{2220}&(f,C_{1311})^{0101}\\
D^2_{2220}&(f,C^1_{1111})^{0001}\\
D^1_{2202}&(f,C_{1113})^{0011}\\
D^2_{2202}&(f,C_{1311})^{0110}\\
D^1_{2022}&(f,C_{1113})^{0101}\\
D^2_{2022}&(f,C^1_{1111})^{0100}\\
D^1_{0222}&(f,C_{1113})^{1001}\\
D^2_{0222}&(f,C_{1311})^{1100}\\\hline
\end{array}\end{array}\nonumber\end{equation*}\end{center}
\begin{center} Degree 5
$$
\begin{array}{|c|c|}\hline
\mbox{Symbol}&\mbox{Transvectant}\\\hline
E_{1111}&(f,D_{2200})^{1100}\\\hline
E^1_{3111}&(f,D_{2200})^{0100}\\
E^2_{3111}&(f,D_{2202}^1)^{0101}\\
E^3_{3111}&(f,D_{2022}^2)^{0011}\\
E^1_{1311}&(f,D_{2200})^{1000}\\
E^2_{1311}&(f,D_{0202})^{0001}\\
E^3_{1311}&(f,D_{0220})^{0010}\\
E^1_{1131}&(f,D_{0222}^1)^{0101}\\
E^2_{1131}&(f,D_{2022}^2)^{1001}\\
E^3_{1131}&(f,D_{2020})^{1000}\\
E^1_{1113}&(f,D_{2022}^1)^{1010}\\
E^2_{1113}&(f,D_{2022}^2)^{1010}\\
E^3_{1113}&(f,D_{0004})^{0001}\\\hline
\end{array}
\nonumber
$$
\end{center}
\begin{center}Degree 6
\begin{equation*}\begin{array}{cc}\begin{array}{|c|c|}\hline
\mbox{Symbol}&\mbox{Transvectant}\\\hline
F_{0000}&(f,E_{1111})^{1111}\\\hline
F_{2200}&(f,E^1_{3111})^{1011}\\
F_{2020}&(f,E_{1111})^{0101}\\
F_{2002}&(f,E^1_{1113})^{0111}\\
F_{0220}&(f,E^1_{1311})^{1101}\\
F_{0202}&(f,E^3_{1113})^{1011}\\
F_{0022}&(f,E^1_{1113})^{1101}\\\hline
F_{2220}^1&(f,E^1_{1311})^{0101}\\
F_{2220}^2&(f,E^2_{1311})^{0101}\\
F_{2202}^1&(f,E^2_{3111})^{1010}\\
F_{2202}^2&(f,E^3_{3111})^{1010}\\
F_{2022}^1&(f,E^1_{1113})^{0101}\\
F_{2022}^2&(f,E^2_{1113})^{0101}\\
F_{0222}^1&(f,E^1_{1131})^{1010}\\
F_{0222}^2&(f,E^2_{1131})^{1010}\\\hline\end{array}&
\begin{array}{|c|c|}\hline
\mbox{Symbol}&\mbox{Transvectant}\\\hline
F_{4200}&(f,E^1_{3111})^{0011}\\
F_{4020}&(f,E^2_{3111})^{0101}\\
F_{4002}&(f,E^2_{3111})^{0110}\\
F_{0420}&(f,E^3_{1311})^{1001}\\
F_{0402}&(f,E^2_{1311})^{1010}\\
F_{0042}&(f,E^1_{1131})^{1100}\\
F_{2400}&(f,E^1_{1311})^{0011}\\
F_{2040}&(f,E^1_{1131})^{0101}\\
F_{2004}&(f,E^1_{1113})^{0110}\\
F_{0240}&(f,E^1_{1131})^{1001}\\
F_{0204}&(f,E^1_{1113})^{1010}\\
F_{0024}&(f,E^1_{1113})^{1100}\\\hline
\end{array}\end{array}\nonumber\end{equation*}
\end{center}
\begin{center}Degree 7
\begin{equation*}\begin{array}{cc}\begin{array}{|c|c|}\hline
\mbox{Symbol}&\mbox{Transvectant}\\\hline
G^1_{3111}&(f,F_{2200})^{0100}\\
G^2_{3111}&(f,F_{4002})^{1001}\\
G^3_{3111}&(f,F_{2202}^1)^{0101}\\
G^1_{1311}&(f,F_{0402})^{0101}\\
G^2_{1311}&(f,F_{2200})^{1000}\\
G^3_{1311}&(f,F_{0202})^{0001}\\
G^1_{1131}&(f,F_{0222}^1)^{0101}\\
G^2_{1131}&(f,F_{0222}^2)^{0101}\\
G^3_{1131}&(f,F_{2040})^{1010}\\
G^1_{1113}&(f,F_{2022}^1)^{1010}\\
G^2_{1113}&(f,F_{2022}^2)^{1010}\\
G^3_{1113}&(f,F_{0202})^{0100}
\\\hline\end{array}&
\begin{array}{|c|c|}\hline
\mbox{Symbol}&\mbox{Transvectant}\\\hline
G_{5111}&(f,F_{4002})^{0001}\\
G_{1511}&(f,F_{0402})^{0001}\\
G_{1151}&(f,F_{2040})^{1000}\\
G_{1115}&(f,F_{0024})^{0010}\\\hline
G_{3311}&(f,F_{2400})^{0100}\\
G_{3131}&(f,F_{2022}^2)^{0001}\\
G_{3113}&(f,F_{4002})^{1000}\\
G_{1331}&(f,F_{0240})^{0010}\\
G_{1313}&(f,F_{0402})^{0100}\\
G_{1133}&(f,F_{2022}^2)^{1000}\\\hline
\end{array}\end{array}\nonumber\nonumber
\end{equation*}
\end{center}
\begin{center} Degree 8
\begin{equation*}\begin{array}{cc}\begin{array}{|c|c|}\hline
\mbox{Symbol}&\mbox{Transvectant}\\\hline
H_{4000}&(f,G_{5111})^{1111}\\
H_{0400}&(f,G_{1311}^1)^{1011}\\
H_{0040}&(f,G_{1151})^{1111}\\
H_{0004}&(f,G_{1113}^3)^{1110}
\\\hline
H_{2220}^1&(f,G_{1311}^1)^{0101}\\
H_{2220}^2&(f,G_{1311}^2)^{0101}\\
H_{2202}^1&(f,G_{3111}^3)^{1010}\\
H_{2202}^2&(f,G_{1113}^2)^{0011}\\
H_{2022}^1&(f,G_{1113}^1)^{0101}\\
H_{2022}^2&(f,G_{1113}^2)^{0101}\\
H_{0222}^1&(f,G_{1131}^1)^{1010}\\
H^2_{0222}&(f,G_{1131}^2)^{1010}\\\hline\end{array}&
\begin{array}{|c|c|}\hline
\mbox{Symbol}&\mbox{Transvectant}\\\hline
H_{4200}&(f,G_{5111})^{1011}\\
H_{4020}&(f,G_{5111})^{1101}\\
H_{4002}&(f,G_{5111})^{1110}\\
H_{0420}&(f,G_{1311}^1)^{1001}\\
H_{0402}&(f,G_{1313})^{1011}\\
H_{0042}&(f,G_{1151})^{1110}\\
H_{2400}&(f,G_{1311}^1)^{0011}\\
H_{2040}&(f,G_{1151})^{0111}\\
H_{2004}&(f,G_{1113}^1)^{0110}\\
H_{0240}&(f,G_{1151})^{1011}\\
H_{0204}&(f,G_{1113}^1)^{1010}\\
H_{0024}&(f,G_{1113}^1)^{1100}\\\hline
\end{array}\end{array}\nonumber
\end{equation*}\end{center}
\begin{center} Degree 9
\begin{equation*}\begin{array}{cc}\begin{array}{|c|c|}\hline
\mbox{Symbol}&\mbox{Transvectant}\\\hline
I_{3111}&(f,H_{4020})^{1010}\\
I_{1311}&(f,H_{2220}^1)^{1010}\\
I_{1131}&(f,H_{0240})^{0110}\\
I_{1113}&(f,H_{2004})^{1001}\\\hline
I_{5111}^1&(f,H_{4020})^{0010}\\
I_{5111}^2&(f,H_{4002})^{0001}\\
I_{1511}^1&(f,H_{0402})^{0001}\\
I_{1511}^2&(f,H_{2400})^{1000}\\
I_{1151}^1&(f,H_{0240})^{0100}\\
I_{1151}^2&(f,H_{0042})^{0001}\\
I_{1115}^1&(f,H_{2004})^{1000}\\
I_{1115}^2&(f,H_{0024})^{0010}\\\hline
\end{array}&
\begin{array}{|c|c|}\hline
\mbox{Symbol}&\mbox{Transvectant}\\\hline
I_{3311}^1&(f,H_{2220}^1)^{0010}\\
I_{3311}^2&(f,H_{2220}^2)^{0010}\\
I_{3131}^1&(f,H_{4020})^{1000}\\
I_{3131}^2&(f,H_{2220}^1)^{0100}\\
I_{3113}^1&(f,H_{2004})^{0001}\\
I_{3113}^2&(f,H_{2022}^1)^{0010}\\
I_{1331}^1&(f,H_{0240})^{0010}\\
I_{1331}^2&(f,H_{2220}^1)^{1000}\\
I_{1313}^1&(f,H_{0204})^{0001}\\
I_{1313}^2&(f,H_{0222}^1)^{0010}\\
I_{1133}^1&(f,H_{0024})^{0001}\\
I_{1133}^2&(f,H_{0222}^1)^{0100}\\\hline
\end{array}\end{array}\nonumber
\end{equation*}\end{center}

\begin{equation*}\begin{array}{cc}\mbox{Degree 10}&\\\begin{array}{|c|c|}\hline
\mbox{Symbol}&\mbox{Transvectant}\\\hline
J_{4200}&(f,I_{5111}^1)^{1011}\\
J_{4020}&(f,I_{5111}^1)^{1101}\\
J_{4002}&(f,I_{3113}^1)^{0111}\\
J_{0420}&(f,I_{1331}^1)^{1011}\\
J_{0402}&(f,I_{1511}^1)^{1110}\\
J_{0042}&(f,I_{1133}^1)^{1101}\\
J_{2400}&(f,I_{1511}^1)^{0111}\\
J_{2040}&(f,I_{3131}^1)^{1101}\\
J_{2004}&(f,I_{3113}^1)^{1110}\\
J_{0240}&(f,I_{1331}^1)^{1101}\\
J_{0204}&(f,I_{1115}^1)^{1011}\\
J_{0024}&(f,I_{1115}^1)^{1101}\\\hline
\end{array}&\begin{array}{c}\mbox{Degree 11}\\
\begin{array}{|c|c|}\hline
\mbox{Symbol}&\mbox{Transvectant}\\\hline
K_{3311}&(f,J_{4200})^{1000}\\
K_{3131}&(f,J_{4020})^{1000}\\
K_{3113}&(f,J_{4002})^{1000}\\
K_{1331}&(f,J_{0420})^{0100}\\
K_{1313}&(f,J_{0402})^{0100}\\
K_{1133}&(f,J_{0042})^{0010}\\\hline
K_{5111}&(f,J_{4200})^{0100}\\
K_{1511}&(f,J_{2400})^{1000}\\
K_{1151}&(f,J_{2040})^{1000}\\
K_{1115}&(f,J_{2004})^{1000}\\\hline
\end{array}\\
\end{array}
\end{array}
\nonumber
\end{equation*}
\begin{center} Degree 12
$$
\begin{array}{|c|c|}\hline
\mbox{Symbol}&\mbox{Transvectant}\\\hline
L_{6000}&(f,K_{5111})^{0111}\\
L_{0600}&(f,K_{1511})^{1011}\\
L_{0060}&(f,K_{1151})^{1101}\\
L_{0006}&(f,K_{1115})^{1110}\\
\hline
\end{array}                                                                        $$
\end{center}

\normalsize

\newpage
\section{\label{appsyz}Syzygies}

The method of associated forms presented in Section \ref{ratcov}
gives all the syzygies up to degree 5. In degree 6, 
one can check that, for example, the following two syzygies
\begin{eqnarray*}
D_{0000}^2f^2+2D_{0000}^1f^2-\case32B_{2200}B_{0022}B_{0000}&\\
+\case32B_{2020} B_{0202}B_{0000} -\case92D_{2200}B_{0022}&\\
-4(C_{1111}^2)^2-\case92D_{0022}B_{2200}-8C_{1111}^1C_{1111}^2&\\
+\case92D_{0220}B_{2002}+\case92D_{2002}B_{0220}&=0
\end{eqnarray*}
\begin{eqnarray*} 
(C_{1111}^1)^2+\case34D_{0000}^1f^2-\case98B_{2200}B_{0022}B_{0000}&\\
+\case98B_{2020}B_{0202}B_{0000} -\case94D_{2200}B_{0022}&\\
+\case98D_{0202}B_{2020}-2(C_{1111}^2)^2-\case94D_{0022}B_{2200}&\\
-2C_{1111}C_{1111}-\case32D_{2020}B_{0202}&\\
+\case98D_{0220}B_{2002}+\case98D_{2002}B_{0220}&=0              
\end{eqnarray*}
cannot be obtained by this method.

The second order syzygies arise in degree 7 and type
$(5333)$.

\section{\label{invcov}Invariants of the covariants}

As already mentioned, the next step in the study of the four qubit
system should be to find geometrical interpretations of the simplest
covariants. Partial information is already available, as
some of the fundamental covariants are of known types, for which
the invariant theory is reasonably well understood.
Among these, the most important ones are certainly the six
biquadratic forms $b_{uv}$. A double binary form $g({\bf u},{\bf v})$
of bidegree $(m,n)$
is usually interpreted as defining a space curve, lying of the 
projective quadric
$XT-YZ=0$, which may be parametrized by 
\begin{equation}
\left(\matrix{X & Y \cr Z & T}\right)
=
\left(\matrix{u_2v_2 & - u_2v_1\cr -u_1v_2 & u_1v_1}\right)\,,
\end{equation} 
so that ${\bf u}$ and ${\bf v}$ parametrize the rectilinear generatrices.
Such a curve has generically $m$ intersections with the generatrices
of one system, and $n$ with generatrices of the other one.
The covariants $b_{uv}$ can therefore be interpreted as six space quartics.
It turns out that these have the same discriminant, which is proportional
to the hyperdeterminant used in \cite{Mi}. The nonvanishing of the
hyperdeterminant is therefore the condition for these curves to be elliptic.
On the normal forms of  \cite{Ver}, it is easy to check that one has
three isomorphisms $(b_{xy})\simeq (b_{zt})$, $(b_{xz})\simeq (b_{yt})$
and $(b_{xt})\simeq (b_{yz})$. Geometric
interpretations of the  complete system of covariants
of such forms of bidegree $(2,2)$ (which is due to Peano \cite{Pea})
are available in the works of Kasner \cite{Kas} and Turnbull \cite{Turn}.

Other double binary forms of the same type occur as
covariants in degrees 4 and 6.

There are also some simple binary forms among the covariants.
Quartics, whose invariant theory is completely understood,
occur in degree 4 and 8. We have already used the fact that the
common discriminant of the quartics in degree 4 was the
hyperdeterminant \cite{LT1}. The fundamental covariants in degree 12,
the highest degree, are four binary sextics, whose invariant
of degree 2 is again the hyperdeterminant, as can be
seen from the normal forms
\begin{eqnarray}
L_{6000}&=&144\,V(a^2,b^2,c^2,d^2)(x_1^4-x_2^4)x_1x_2\\
L_{0600}&=&96\,V(a^2,b^2,c^2,d^2)(y_1^4-y_2^4)y_1y_2\\
L_{0060}&=&288\,V(a^2,b^2,c^2,d^2)(z_1^4-z_2^4)z_1z_2\\
L_{0006}&=&-96\,V(a^2,b^2,c^2,d^2)(t_1^4-t_2^4)t_1t_2
\end{eqnarray}
where $V$ denotes the Vandermonde determinant.
One can
see from this example that the normal forms of  \cite{Ver}
are simple enough to allow the complete calculation of
all the fundamental covariants. It is also interesting
to observe that under the specialization $G_{abcd}$ of
\cite{Ver}, the covariants $b_{uv}$ come directly in Kasner's normal form      
\begin{equation}
k (u_1^2v_1^2+u_2^2v_2^2)+ l(u_1^2v_2^2+u_2^2v_1^2)+4m\, u_1u_2v_1v_2
\end{equation}
and that the three quadrilinear covariants occuring in degrees 3 and 5
are also in normal form. Indeed,
\begin{eqnarray*}
C_{1111}^1=& \case{a_1 +  d_1}2 x_1 y_1 z_1 t_1 + \case{ a_1 -  d_1}2 x_1 y_1 z_2 t_2\\
    & + \case{ b_1 + c_1 }2x_1 y_2 z_1 t_2  + \case{ b_1 -  c_1}2 x_1 y_2 z_2 t_1\\
& + \case{ b_1 -c_1}2 x_2 y_1 z_1 t_2 + \case{ b_1 +  c_1}2 x_2 y_1 z_2 t_1\\
&     + \case{ a_1 -  d_1}2 x_2 y_2 z_1 t_1 + \case{a_1 + d_1}2 x_2 y_2 z_2 t_2 \,,
\end{eqnarray*}  
where
\begin{eqnarray*}
a_1 &=& 3a^3-ad^2-ab^2-ac^2\\
b_1 &=& -bc^2-ba^2-bd^2+3b^3\\
c_1 &=& -b^2c+3c^3-ca^2-cd^2\\
d_1 &=& -a^2d+3d^3-db^2-dc^2\,,
\end{eqnarray*} 
and 
\begin{eqnarray*}
C_{1111}^2=& \case{a_2 +  d_2}2 x_1 y_1 z_1 t_1 + \case{ a_2 -  d_2}2 x_1 y_1 z_2 t_2\\
   &  + \case{ b_2 + c_2 }2x_1 y_2 z_1 t_2 + \case{ b_2 -  c_2}2 x_1 y_2 z_2 t_1\\
   &  + \case{ b_2 -c_2}2 x_2 y_1 z_1 t_2 + \case{ b_2 +  c_2}2 x_2 y_1 z_2 t_1\\
   &  + \case{ a_2-  d_2}2 x_2 y_2 z_1 t_1 + \case{a_2 + d_2}2 x_2 y_2 z_2 t_2\,, 
\end{eqnarray*} 
where
\begin{eqnarray*}
a_2 &=& 2ab^2+2ac^2+2ad^2+6dbc\\
b_2 &=& 2ba^2+2bd^2+2bc^2+6cad\\
 c_2& =& 2b^2c+6bad+2ca^2+2cd^2\\
 d_2& =& 2a^2d+6abc+2db^2+2dc^2\,
\end{eqnarray*}
and finally
\begin{eqnarray*}
E_{1111}=& \case{a_3 +  d_3}2 x_1 y_1 z_1 t_1 + \case{ a_3 -  d_3}2 x_1 y_1 z_2 t_2\\
    & + \case{ b_3 + c_3 }2x_1 y_2 z_1 t_2 + \case{ b_3 -  c_3}2 x_1 y_2 z_2 t_1\\
    & + \case{ b_3 -c_3}2 x_2 y_1 z_1 t_2 + \case{ b_3 +  c_3}2 x_2 y_1 z_2 t_1\\
   &  + \case{ a_3 -  d_3}2 x_2 y_2 z_1 t_1 + \case{a_3 + d_3}2 x_2 y_2 z_2 t_2\,,
\end{eqnarray*}
where
\begin{eqnarray*}
 a_3& =& 8(-a^3d^2-c^2a^3-b^2a^3+ac^2d^2+ab^2c^2+ab^2d^2)\\
b_3 &=& 8(-b^3c^2+bc^2d^2+bc^2a^2-b^3d^2+ba^2d^2-b^3a^2)\\
c_3 &=& 8(ca^2d^2+a^2b^2c+cb^2d^2-b^2c^3-c^3d^2-c^3a^2)\\
d_3&=& 8(db^2c^2+a^2b^2d+a^2c^2d-a^2d^3-c^2d^3-b^2d^3)
\end{eqnarray*}

On another hand, it seems that nothing is known about quadruple
binary forms of multidegree $(3,1,1,1)$, and the first thing
to be done now is probably to set up a convenient geometric representation
of such forms.

\end{document}